# Free standing membranes to study the optical properties of anodic TiO$_2$ nanotube layers

by Gihoon Cha,[1,†] Patrik Schmuki,[1,2]* and Marco Altomare,[1,†]*

[1] Gihoon Cha, Prof. Dr. Patrik Schmuki, Dr. Marco Altomare

Department of Materials Science and Engineering WW4-LKO, University of Erlangen-Nuremberg,

Martensstrasse 7, Erlangen, D-91058, Germany

[2] Prof. Dr. Patrik Schmuki

Department of Chemistry, King Abdulaziz University, Jeddah, Saudi Arabia

* Corresponding Authors; E-mail: schmuki@ww.uni-erlangen.de; marco.altomare@fau.de

[†] Both authors contributed equally

**ABSTRACT**

In the present work we investigate various optical properties (such as light absorption and reflectance) of anodic $TiO_2$ nanotubes layers directly transferred as self-standing membranes onto quartz substrates. This allows investigation in a transmission geometry which provides significantly more reliable data than measurements on the metallic Ti substrate. Light transmission and reflectance measurements were carried out for layers of thicknesses varying from 1.8 to 50 μm, and the layers were investigated in their amorphous and crystalline form. A series of wavelength-dependent light attenuation coefficients are extrapolated and found to match the photocurrent *vs.* irradiation wavelength behavior. However, a feature specific to anodic nanotubes is that their intrinsic carbon content causes a sub-bandgap response that is proportional to the carbon contamination content in the $TiO_2$ nanotubes. Overall the extracted data provide valuable basis and understanding for the design of photo-electrochemical devices based on $TiO_2$ nanotubes.

**Introduction**

Among the different metal oxide semiconductors, titanium dioxide ($TiO_2$) is by far the most widely investigated material, owing to its low cost, large availability, high (photo-)chemical stability, and, most importantly, to the proper position of valence and conduction bands that make $TiO_2$ a most promising candidate for several catalytic and photo-electrochemical applications.[1–3]

$TiO_2$, having in its anatase and rutile forms bandgap of ~ 3.2 and 3.0 eV,[4] respectively, is widely explored as a photocatalyst for water splitting, since valence band holes and conduction band electrons generated upon super-bandgap irradiation are able to oxidize and reduce water into $O_2$ and $H_2$, respectively.[5] In addition, owing to its relatively high electron mobility,[6–8] $TiO_2$ scaffolds have found wide application as anodes for photo-electrochemical (PEC) water splitting,[5,9,10] and for dye-sensitized solar cells (DSSCs).[11,12]

Owing to the extended use in such light driven applications and devices, large efforts have been given to study $TiO_2$ optical properties. On the one hand, $TiO_2$ is explored as a solid state light absorber, and suitable analytical tools for this purpose are measurements of absorbance, transmittance and reflectance.[13–18] On the other hand, since as anticipated $TiO_2$ is often used as an illuminated anode immersed into a suitable electrolyte, photo-electrochemical techniques are employed to determine parameters such as the semiconductor bandgap, the light absorption threshold, and the wavelength dependence of the generated photocurrent.

In this work we investigate by different techniques the optical properties of anodic $TiO_2$ nanotube layers, the growth of which is based on the simple anodization of a piece of Ti metal under self-organizing electrochemical conditions.[19,20] This fabrication approach has gained large attention in the last decade as it is a straightforward and versatile nanostructuring technique that can lead to several controlled geometries of self-organized structures by adjusting the electrochemical conditions, and it allows for the growth of oxide films that are back-contacted, and therefore ready to be used, *e.g.*, as an electrode in photo-electrochemical applications. In addition, the obtained TiO$_2$ architectures show a series of functional features that are, namely, the large specific surface area ascribed to the nanostructured morphology,[19] their one-dimensionality that leads to preferential percolation pathways for photogenerated electrons, the orthogonal separation of charge carriers and the reduced hole diffusion pathway towards the TiO$_2$ surface.[21–23]

Noteworthy, while a number of measurements on some optical properties of TiO$_2$ nanotube layers have been reported in the literature, the possibly most important geometry, *i.e.*, the measurement of light transmittance is so far unexplored. The reason for this is that TiO$_2$ nanotube layers are typically grown on Ti metal foils and therefore cannot be used for light transmission measurements.

In this work we illustrate how by an adequate anodic growth sequence followed by a controlled chemical etching, TiO$_2$ nanotube layers can be detached from the Ti metal substrate and transferred as self-standing membranes onto quartz slides (without the use of an adhesive layer). We show that the lift-off and transfer strategy can be used to form tube membranes as thin as of ~ 1.8 μm. A further key of this approach is that we avoid the use of intermediate layer deposited between the tube arrays and the quartz slide, which would interfere with the optical measurements. Note that in all prior work using membranes, *e.g.*, for solar cells, films of TiO$_2$ nanoparticles are typically deposited onto the quartz to improve the membrane adhesion and its mechanical stability.[24–26]

We report for our membranes a complete set of light transmission and reflectance data, in relation to the irradiation wavelength, the TiO$_2$ nanotube layer thickness and their amorphous or crystalline nature. Particularly, we extract from the transmission results a series of wavelength-dependent light attenuation coefficients, and show that these can be used as a constant specific of

anatase-TiO$_2$-tube films that describes how monochromatic light is attenuated when passing through the nanotube layer. We explore also different annealing conditions for the tube layers and show that their light absorption and photo-electrochemical properties in the visible range are related to the carbon content in the anodic films.

**Experimental**

Prior to anodization, Ti sheets with a thickness of 0.125 mm (99.6 % purity, Advent Materials, UK) were cleaned in acetone, ethanol, and distilled water (15 min for each treatment) in an ultrasonic bath, and then were dried in a N$_2$ stream.

The TiO$_2$ nanotube layers were grown by electrochemical anodization in a glycol ethylene-based electrolyte containing 0.15 M NH$_4$F and 3 vol% H$_2$O (all reagents were provided by Sigma Aldrich). The electrochemical experiments were carried out always in a two-electrode electrochemical cell, with the Ti foil and a Pt gauze being the working and the counter electrodes, respectively. The applied potential was provided by a LAB/SM 1300 DC power supply (ET System). The thickness of the nanotube layers was adjusted by the anodization time. After the anodization, the nanotube films were rinsed with ethanol and dried in a N$_2$ stream. The lift-off procedure for the nanotube membranes, their transfer onto optically transparent quartz slides and their annealing are described in the Results and Discussion Section (and are also illustrated in Scheme 1). The membranes were exposed to thermal treatment using a Rapid Thermal Annealer (Jipelec Jetfirst 100 RTA), with a heating and cooling rate of 30°C min$^{-1}$. The quartz slides were 2.5 cm x 1.5 cm x 0.1 cm and were provided by GVB GmbH.

For morphological characterization, a field-emission scanning electron microscope (Hitachi FE-SEM S4800) was used. The cross-sectional images were obtained by cracking the TiO$_2$ nanotube layers. X-ray diffraction analysis (XRD, X'pert Philips MPD with a Panalytical

X'celerator detector) using graphite monochromized Cu Kα radiation (wavelength 1.54056 Å) was used for determining the crystallographic composition of the samples. Energy-dispersive X-ray spectroscopy (EDAX Genesis, fitted to SEM chamber) was also used for the chemical analysis of the anodic layers.

The light transmission measurements were carried out by irradiating the nanotube layers with monochromatic light provided by four different lasers (266 nm, FQCW266-50, CryLas, $I_0$ = 8 mW; 325 nm, IK3552R-G, KIMMON $I_0$ = 23 mW; 405 nm, MDL-III-405, $I_0$ = 19.2 mW; 473 nm, MBL-III-473, $I_0$ = 10 mW). The intensity of the monochromatic light passing through the TiO2 nanotube layers was measured using a 1830-C (Newport) calibrated power meter. Diffuse reflectance *spectra* (DRS) in the 375-600 nm range were measured using a LAMBDA 950 UV-Vis spectrophotometer (Perkin Elmer, Beaconsfield, UK).

The photocurrent *spectra* were recorded at a constant potential of + 0.5 V (*vs.* Ag/AgCl) provided by a Jaissle IMP83 PC-T-BC potentiostat. A three-electrode electrochemical cell was used, with the anodic film and a Pt gauze being the working and the counter electrodes (a Ag/AgCl was used as reference). The measurements were carried out in $Na_2SO_4$ aqueous solutions (0.1 M) under irradiation provided by a Oriel 6365 150 W Xe-lamp, emitting in the 200-800 nm range. The irradiation wavelength incident to the samples was selected by using a motor driven monochromator (Oriel Cornerstone 130 1/8 m). Incident photon to energy conversion (IPCE) values were calculated using the following equation: $IPCE\% = \frac{1240 \times J_{ph}}{\lambda \times I_{light}}$, where $J_{ph}$ is the measured photocurrent density (mA cm$^{-2}$), λ is the incident light wavelength (nm), and $I_{light}$ is the intensity of the light source at a specific wavelength (mW cm$^{-2}$).

The nanotube layers used for the photo-electrochemical measurements were grown from Ti foils (up to a thickness of ~ 10 μm) under the same experimental conditions adopted for fabricating the membranes. These nanotube layers were investigated both as amorphous and crystalline. In the case of crystalline films, the nanotube layers were converted to anatse phase by annealing at 500°C (in the RTA), in air, for 1 or 4 h. The compact anodic films were produced in aqueous electrolytes (1 M $H_2SO_4$), at 20 V for 20 min, and were annealed under the same conditions used for the nanotube layers.

**Results and Discussion**

Scheme 1 illustrates the procedure we used to produce the $TiO_2$ nanotube membranes studied in this work. To fabricate membranes of highly ordered nanotubes, we employed a double-anodization approach that previously was explored particularly for the growth of anodic alumina nanopores layers.[27] For this we perform a first anodization of the Ti foil at 60 V for 20 min to grow a *ca.* 10 µm-thick tube layer that after anodization is removed by sonication. The removal of such a tube layer leaves on the surface of the Ti metal substrate an array of dimples, *i.e.*, a replica of the tube bottom morphology.

The Ti surface is then anodized for different times, from 4 min up to 14 h, to grow nanotube layers with a thickness of *ca.* 1.8-50 µm. During this anodization step the dimples on the Ti surface act as preferential initiation site for the anodic growth, leading to highly-ordered and aligned structures.[28] Subsequently the sample is exposed to an annealing in air at low temperature (250°C) to dehydrate and improve the chemical stability of the tube layer. Then the sample is anodized once more to grow a tube layer underneath the annealed layers. After anodization the samples are treated in a 30% $H_2O_2$ aqueous solution at room temperature for *ca.* 3 h. Owing to the higher chemical stability of the upper anodic layer, this treatment leads to the chemical dissolution of the lower tube layer, so that the upper one is detached from the substrate as entire self-standing $TiO_2$ nanotube membrane.[12,29,30]

After detachment, the tube membranes of different thickness are transferred onto quartz slides in a tube-top-down configuration, covered with a porous ceramic block and left overnight in air at room temperature for drying. When the porous block is removed, the membranes stick well to the quartz glasses and have good mechanical stability even after annealing.

The $TiO_2$ nanotube membranes were explored in view of their optical properties both in amorphous (as-formed) and crystalline state. The crystallization was generally carried out by exposing the tube membrane deposited on quartz to a thermal treatment in air, at 500°C for 1h. In some cases, the thermal treatment implied also a second annealing under the same conditions for another 3 h. In any case, another quartz slide was placed on the membranes before annealing (*i.e.*, the membrane was sandwiched between two quartz slides) to avoid cracking and curling effects.

The optical picture in Scheme 1 shows the as-formed $TiO_2$ nanotubes membranes successfully transferred onto quartz slides. From right to left, the thickness of the membranes increases, from *ca.* 1.8 up to 50 µm.

The SEM images in Fig. 1 (see also in Fig. S1) show the morphology of the $TiO_2$ nanotubes used in this work that has an average inner diameter of *ca.* 60 nm. From Fig. 1 (a,c,e) (bottom view) it is evident that the membranes are open at the bottom, that is, during the prolonged lift-off step in aqueous $H_2O_2$, not only the lower anodic layer but also the bottom of the upper tube layer is chemically etched, resulting in a flow-through tube morphology.[31] Moreover, from the cross-sectional view (Fig. 1 (b,d,f)) one can clearly see that in general the anodic layers grow as arrays of highly aligned and ordered nanotubular structures, and their thickness is precisely adjusted (by the anodization time) over a quite large thickness range.

As reported in the literature, as-formed anodic $TiO_2$ nanotube layers are amorphous, and this is confirmed by the XRD data in Fig. 2 (a). XRD also showed their crystallization to anatase $TiO_2$ by a thermal treatment in air at 500°C – the absence of the Ti foil underneath the tube layers avoid the formation of Ti oxide from the Ti metal.[32–34] It is also evident that the 4 h-long annealing, compared to a thermal treatment of 1 h, leads to tube membranes with higher degree of crystallinity as the intensity of the main anatase peak (at *ca.* 25.3°) increases with the annealing time. The XRD data in Fig. 2 (b) show that the intensity of the anatase signal increases with the thickness of the tube membranes (annealed in air, 500°C), which is ascribed to the larger content of crystalline oxide in thicker tube layers.

The amorphous and anatase $TiO_2$ membranes of different thicknesses were studied in view of their light transmission properties. For this we used four different lasers, emitting at 266, 325, 405 and 473 nm. The laser light was directed onto the membranes deposited on the quartz slides, and a calibrated power meter placed behind the oxide/quartz was used to determine the intensity of the transmitted light – thus, these measurements provide a direct assessment of the extent by which the intensity of the incident monochromatic light is reduced as the light passes through the $TiO_2$ nanotube film (due to absorption, scattering). We illustrate below that these measurements can be used to assess both qualitatively and quantitatively the dependence of the optical properties of $TiO_2$ nanotube layers on their thickness and amorphous or crystalline nature, and also on the irradiation wavelength.

The light transmission data measured at four different wavelengths are compiled in Fig. 3, and are plotted as $I/I_0$ *vs.* the thickness of the tube membranes. For each wavelength, $I$ is the measured intensity of light passing through the different nanotube films, while $I_0$ is the full light intensity emitted by the different lasers after passing through a plain quartz slide (*i.e.*, regardless of the irradiation wavelength, the quartz slides do not significantly attenuate the laser light intensity). Therefore, $I/I_0$ is directly proportional to the film transmittance, and inversely proportional to both the absorbance of the films and light scattering *phenomena* – in other words, the larger the absorbance (or the scattering) of the tube film the lower the intensity of transmitted light.

A first evident result is that for all the irradiation wavelengths, the intensity of transmitted light drops steeply with increasing the thickness of the tube films. A second aspect is that amorphous films always show lower light transmission compared to crystalline tubes of the same thickness.

In the UV region (266 and 325 nm, *i.e.*, super-bandgap irradiation) the amorphous or crystalline nature of the tubes affects their light transmission only for relatively thin layers, *e.g.*, < 5 µm. This is particularly clear for 1.8 µm-thick tube layers that transmit only ~ 1% of the laser full emission intensity when they are amorphous, while their light transmission increases up to ~ 11% upon conversion into anatase $TiO_2$ – both amorphous and crystalline films thicker than 5 µm transmit almost no light (*i.e.*, $I/I_0$ ~ 0%).

A similar qualitative trend can be seen under visible light irradiation (405 and 473 nm). However, as expected for sub-bandgap illumination of both amorphous and crystalline layers, the light transmission is much higher than that measured in the UV range. In addition, the intensity of transmitted light for thick films (even up to 50 µm) significantly increases upon their conversion into anatase $TiO_2$, and the longer the irradiation wavelength the larger the relative increase of $I/I_0$.

Overall these results confirm that upon crystallization into anatase $TiO_2$, the membranes show a low transmission of UV light compared to visible light, in line with the bandgap of anatase $TiO_2$ (which corresponds to a light absorption threshold of *ca.* 390-410 nm). Note in fact that for 1.8 µm-thick anatase layer, the light transmission at 266 and 325 nm (ultra-bandgap irradiation) is almost constant at ~ 12 %, it increases at 405 nm (close to the absorption threshold) up to ~ 18%, and it increases even more up to ~ 26% at 473 nm (sub-bandgap irradiation) – in other words, the different light transmission measured under UV or visible light irradiation is ascribed to light absorption (as opposed to *e.g.* scattering). Nevertheless, these features can be observed only for thin layers since, as the intensity of transmitted light significantly drops with increasing the thickness of the membranes. This saturation trend of transmitted light can be ascribed to light scattering effects that become predominant in thicker film, and are more evident for amorphous layers regardless of the irradiation wavelength. On the contrary, for thick anatase layers the larger the irradiation wavelength the less pronounced the saturation effect, that is, while for 28-50 μm-thick anatase layers the light transmission is almost 0% under super-bandgap irradiation, it increases to ~ 1.5% at 405 nm, and up to ~ 4% at 473 nm.

The data compiled in Fig. 3 can also be used to estimate a wavelength dependent light absorption coefficient for the $TiO_2$ nanotube layers. The light absorption of a solid state system (*i.e.*, the tube layers) can be described, as for a light absorber in a liquid solution, using the Lambert-Beer law that can be expressed as in equation (1):

$$A = \varepsilon_{TiO_2,\lambda}\, t \qquad (1)$$

Where *A* is the light absorption at a specific wavelength ($\lambda$), $\varepsilon_{TiO_2,\lambda}$ is the wavelength-dependent light attenuation (absorption) coefficient (typical constant of the light absorbing *medium*), and *t* is the thickness of the tube membrane – as the light attenuation increases with the optical path length through the light absorber it is reasonable to assume that the thicker the tube layer the larger the light absorption.[35,36] Noteworthy, the light absorption coefficient $\varepsilon_{TiO_2,\lambda}$ in equation (1) not only relates to a certain irradiation wavelength, but also depends on the optical density of the light absorber, *i.e.*, the effective optical density of the $TiO_2$ nanotube layers, that is assumed to be constant (as ascribed to the uniform tube morphology). The Lambert-Beer law also allows for relating the light absorption (*A*, *i.e.*, Absorbance) to the light transmittance (*T*, *i.e.*, Transmittance) as described by equation (2) and (3):

$$T = \frac{I}{I_0} \qquad (2)$$

$$\frac{I}{I_0} = 10^{-A} \qquad (3)$$

As described above for the light transmission experiments, $I$ and $I_0$ are the intensity of the transmitted light measured for the different nanotube layers and the full intensity emitted by the laser, respectively. By properly combining equations (1), (2) and (3), equation (4) can be obtained:

$$\frac{I}{I_0} = 10^{-\varepsilon_{TiO_2,\lambda}\, t} \qquad (4)$$

By rearranging equation (4), equation (5) can be obtained:

$$-\log_{10}\frac{I}{I_0} = \varepsilon_{TiO_2,\lambda}\, t \qquad (5)$$

Therefore, according to equation (5), a plot of $A = -\log_{10}\frac{I}{I_0}$ vs. the thickness of the tube membranes is ideally a linear function (as explained below, for $A < 2.5$) from the slope of which the wavelength-dependent light absorption coefficient of $TiO_2$ nanotube layers (i.e., $\varepsilon_{TiO_2,\lambda}$) can be extrapolated.

The light transmission data, i.e., $-\log_{10}\frac{I}{I_0}$, for anatase $TiO_2$ nanotube membranes are plotted for the different irradiation wavelengths in Fig. 4 (a) as described in equation (5), that is, vs. the layer thickness ($t$). Data points that lead to $A$ values larger than 2.5 are not taken into account (for the sake of avoiding instrumental deviation from the Lambert-Beer law). The reason is that for $A > 2.5$, $T\,(\%) = \frac{I}{I_0}100\% < \sim 0.3\%$, that is, for a calculated absorbance of 2.5 the measured intensity of transmitted light is less than 0.3% the full intensity emitted by the laser, which is in the range of instrumental error of the power meter.

For amorphous membranes, $\varepsilon_{TiO_2,\lambda}$ values of ~ 0.17-0.33 $\mu m^{-1}$ are obtained under visible light irradiation – if one assumes that the intensity reduction of the light passing through the films is only associated to light absorption phenomena (i.e., bandgap excitation), these results would in principle indicate a broad band light absorption of the amorphous layers. However, owing to the low light transmission of amorphous layers ($A$ almost always > 2.5 regardless of their thickness), the extrapolated $\varepsilon_{TiO_2,\lambda}$ values represent only a rough estimate.

Nevertheless the as-formed layers show a clearly low light transmission under both UV and visible light irradiation and, even more, one can see from the optical picture in Scheme 1 that these membranes are colored, *i.e.*, the films interfere with visible light, and the thicker the membrane the darker is its color. Hence, a most likely explanation for this is that the relatively high $\varepsilon_{TiO_2,\lambda}$ values along with the color of the amorphous layers are ascribed to the intrinsic carbon content of anodic nanotubes (*i.e.*, carbon species up take of the oxide layers during the anodization in the organic-based electrolyte[20,37] – this aspect is described below with more details).

On the contrary, the light transmission data of crystalline tube membranes lead to A values almost always < 2.5 and this allowed for a reliable evaluation of $\varepsilon_{TiO_2,\lambda}$. According to the model and assumptions described above, the linear fitting of the data in Fig. 4 (a) provides the light absorption coefficient $\varepsilon_{TiO_2,\lambda}$ of anatase nanotube membranes at the different wavelengths. The linear portion of data measured under 266 and 325 nm irradiation are clearly steeper than those measured at 405 and 473 nm. Precisely, the $\varepsilon_{TiO_2,\lambda}$ values are found to be 0.39, 0.43, 0.045 and 0.032 μm$^{-1}$ at 266, 325, 405 and 473 nm, respectively.

The different $\varepsilon_{TiO_2,\lambda}$ values are compiled in Fig. 4 (b) where they are plotted as a function of the irradiation wavelength. In the same plot we also show a typical photocurrent (IPCE) *spectrum* measured for a 10 μm-thick layer of anatase TiO$_2$ nanotubes (the measurements were carried out in 0.1 M Na$_2$SO$_4$, under an anodic bias of 0.5 V, in a three-electrode electrochemical cell – see the Experimental Section for additional information).

Interestingly, the wavelength dependence of the calculated $\varepsilon_{TiO_2,\lambda}$ values (*i.e.*, the trend of $\varepsilon_{TiO_2,\lambda}$ *vs.* the irradiation wavelength) clearly matches the experimental IPCE spectrum of the anatase $TiO_2$ nanotube layer. This indicates that the model in equation (5) holds in the case of anatase nanotube layers, and the evident match in Fig. 4 (b) between the calculated $\varepsilon_{TiO_2,\lambda}$ values and the experimental photocurrent spectrum confirms the validity of the assumption that the measured photocurrent is directly proportional to the light absorption coefficient of anatase $TiO_2$ nanotube layers, as expressed in equation (6)

$$J_{ph,\lambda} = \propto \varepsilon_{TiO_2,\lambda} \qquad (6)$$

where $J_{ph,\lambda}$ is the measured photocurrent at a specific wavelength.[38–40] Also, the trend of $\varepsilon_{TiO_2,\lambda}$ *vs.* the irradiation wavelength allows for a qualitative evaluation of the optical bandgap of anatase $TiO_2$ nanotubes, that matches the photo-electrochemically measured bandgap.

From a practical point of view, the extrapolated $\varepsilon_{TiO_2,\lambda}$ values can be used, as well as the extinction coefficient of a light absorber in a liquid phase, as a wavelength dependent light attenuation coefficient, that is, as a constant parameter that in this case is specific for anatase $TiO_2$ nanotube layers and describe how the monochromatic incident light is attenuated (absorbed) when it passes through the semiconductor.

Since the anatase $TiO_2$ membranes are arrays of bottom-opened nanotubes having almost ideally a cylindrical shape (with outer and inner diameters of ~110 and 60 nm, respectively), one can estimate the porosity of the membranes as of ~30%, that is, the $TiO_2$ anatase phase (*i.e.*, the light absorbing element) occupies 70% of the total geometrical volume. Therefore, to take into account the membrane porosity, the $\varepsilon_{TiO_2,\lambda}$ values measured at 265 and 325 nm (being 0.39 and 0.43 $\mu m^{-1}$, respectively) can be corrected (divided) by a factor 0.7, so to obtain attenuation coefficients that virtually represent how light is absorbed by anatase $TiO_2$ in the form of a compact film (*i.e.*, porosity ~0%). Upon applying the correction factor, the $\varepsilon_{TiO_2,\lambda}$ values become 0.56 and 0.61 µm$^{-1}$ (at 265 and 325 nm, respectively) – note that these values are higher than those reported above for the tube membranes, and this is in line with expectation that a denser TiO$_2$ anatase film, owing to a higher optical density, absorbs more light than a porous layer of the same thickness.

The optical properties of the nanotube membranes deposited on quartz substrates were further explored in the visible range by means of diffuse reflectance spectroscopy (Fig. 5). The data in Fig. 5 (a,b) show a clear dependence of the reflectance on the membrane thickness, that is, an evident drop of the reflectance can be seen with increasing the thickness of the nanotube layers, this regardless of their amorphous or crystalline nature.

Besides, an even more relevant aspect is that a dramatic increase of reflectance can be seen after annealing the membranes to anatase TiO$_2$. Particularly, thin nanotube membranes (< 10 µm) converted to anatase show in the entire visible range a reflectance of ~ 100% (this is also visible to the naked eye as these layers becomes transparent after annealing). Thicker layers show also an increase of reflectance upon annealing, although their reflectance never reaches 100% in the whole visible range, *e.g.*, for 10-28 µm-thick layers, the reflectance at 450 nm is of ~ 40-80% (noteworthy, these membranes do not turn transparent upon annealing). A possible explanation for these results is that thick membranes show more pronounced light scattering *phenomena* compared to their thin counterparts (*i.e.*, the thicker the membrane the larger the light scattering).

The increase of reflectance upon air annealing is evident from the data in Fig. 5 (c), where one can see that the reflectance of a 10 µm-thick tube layer dramatically increases by exposing the membrane to a thermal treatment at 500°C in air for 1h. Even more, the reflectance of the anatase films can further increase by a 4 h-long annealing, *i.e.*, by a prolonged thermal treatment, and the increase of reflectance in the visible range is accompanied by an evident color change of the membranes: as shown in the Inset in Fig. 5 (c), while an as-formed layer appears black, it turns pale-yellow after 1 h-long annealing, and becomes white by a thermal treatment at 500°C extended for another 3 h.

Nanotube membranes of different thicknesses and of both amorphous and crystalline nature were characterized by EDX to explore the relation between different annealing treatments and the chemical composition and optical properties of the layers.

The EDX data in Fig. 6 (c) (and also in Fig. S2) show that a carbon content of ~ 10.4 at% is measured for as-formed nanotube layers. For tube layers (of different thicknesses, from 5 to 28 µm) annealed in air, at 500°C for 1 h have a lower carbon content of ~ 2.5-3 at%. Thus, it is evident that the thermal treatment in air (up to 4 h) leads only to a partial removal of the carbon species in the tube layers.

The effects of the carbon remnants and its content in the membranes on the optical properties were clarified by photo-electrochemical measurements of amorphous and differently annealed tube layers (see the Experimental Section for further details). The photocurrent transients (light on/off cycles) in Fig. 7 (a,b) show that, as expected, amorphous $TiO_2$ nanotubes do not show any photocurrent when irradiated at 470 and 500 nm (*i.e.*, owing to the bandgap of $TiO_2$, irradiation with visible light should not induce charge carrier separation in the semiconductor). However, we found that after a 1 h-long air-annealing the anatase tube layers show a clear photocurrent transient. Even more, the photocurrent decreased after annealing the nanotube layer a second time in air for another 3 h. As shown in Fig. 7 (c), this trend was also confirmed by measuring the photocurrent of amorphous and differently annealed (1 and 4 h) tube layers under monochromatic irradiation with a visible light laser emitting at 473 nm (these experiments were performed to avoid possible artifacts of the IPCE setup).

These results overall demonstrate that a relatively short (*e.g.*, 1 h) annealing step in air is on the one hand long enough to convert the amorphous nanotubes into crystalline anatase phase, and on the other hand is too short to establish complete removal of carbon from the tubes. Nevertheless, carbon clearly acts as a visible light sensitizer, that is, either by inducing $TiO_2$ doping or by generating electronic surface states in the bandgap of $TiO_2$,[20] it leads to an evident visible light photo-activity, which is in line with well-established findings on C-doping in $TiO_2$.[41,42] Accordingly, a longer thermal treatment in air leads to a decrease of the carbon content and, as a direct consequence, to a decrease of the visible light photocurrent.

These results, along with reflectance and light transmission results, provide evidence that the high light absorption of anatase tube layers in the visible range can be attributed to the presence of carbon remnants from the electrolyte. Please note that the photocurrent measured for anodic layers grown in non-organic electrolytes (*i.e.*, aqueous $H_2SO_4$) was found to be independent of the annealing time (Fig. 7 (d)), and the reason for this is more likely that the carbon in films grown in water-based electrolyte, which is typically related only to carbon adsorption at the oxide surface ascribed to air-contamination, does not lead to visible light sensitization effects (from EDX analysis the carbon content of as-formed compact $TiO_2$ layer is of ~ 4.5 at%).

**Conclusions**

In this work we illustrate an optimized anodic growth-chemical etching sequence for fabricating self-standing $TiO_2$ nanotube membranes, having a thickness of 1.8 up to 50 μm. The transfer of these membranes onto optically transparent substrates is essential as it allowed for their characterization not only by classical reflectance measurements, but also by light transmission measurements, that to our knowledge are unprecedented for the case of anodically grown $TiO_2$ nanotube layers. For anatase $TiO_2$ nanotube films we extracted from the results of light transmission measurements the light attenuation coefficient, and showed that its dependence on the irradiation wavelength is in line with anatase $TiO_2$ bandgap and nanotube photocurrent *spectra*. We showed also that the photocurrent measured for $TiO_2$ nanotubes under sub-bandgap irradiation is ascribed to the carbon present in the anodic films, the content of which can be adjusted by the annealing conditions.

**Acknowledgements**

The authors would like to acknowledge the ERC, the DFG, and the DFG "Engineering of Advanced Materials" cluster of excellence for financial support. Kiyoung Lee, Xuemei Zhou and Nhat Truong Nguyen are acknowledged for technical help.

**Figure Captions**

**Scheme 1** (a) Schematic of the fabrication of the self-standing $TiO_2$ nanotube membranes. The optical image shows the membranes transferred onto quartz glass (from right to left, their thickness increases as follows: 1.8, 2.5, 5.0, 10.0, 15.0, 28.0 and 50.0 μm).

**Figure 1** SEM images of $TiO_2$ nanotube membranes of different length. (a), (b) 1.8 μm; (c), (d) 10.0 μm; (e), (f) 50.0 μm. SEM images (a), (c) and (e) show the morphology of the bottom of the tube membranes (the inset in (a) shows the membrane top morphology), while images (b), (d) and (f) show their cross-sectional view.

**Figure 2** X-ray diffraction patterns of different $TiO_2$ nanotube membranes: (a) 10 μm-thick membranes, as-formed and annealed in air at 500°C for 1 or 4 h; (b) membranes of different thicknesses (1.8, 10 and 50 μm) annealed in air at 500°C for 1h.

**Figure 3** Light transmission data of as-formed and crystalline (anatase phase) $TiO_2$ nanotube membranes of different thicknesses (1.8-50 μm) measured at different wavelengths: (a) 266 nm; (b) 325 nm; (c) 405 nm; and (d) 473 nm.

**Figure 4** (a) Absorption values calculated from the light transmission data measured at different wavelengths for anatase $TiO_2$ nanotube membranes (the lines represent linear fittings); (b) calculated light attenuation coefficient of anatase $TiO_2$ nanotube membranes and a typical photocurrent trend (of a 10 μm-thick anatase $TiO_2$ nanotube layer) *vs.* the irradiation wavelength.

**Figure 5** Reflectance measurements of: (a) as-formed and (b) crystalline (anatase phase) $TiO_2$ nanotube membranes of different thicknesses (1.8-50 µm); (c) 10 µm-thick $TiO_2$ nanotube membranes as-formed and annealed in air at 500°C for 1 and 4 h. The inset in (c) shows, from left to right, optical images of as-formed, 1 h-annealed and 4 h-annealed 10 µm-thick $TiO_2$ nanotube membranes (annealing was always in air at 500°C).

**Figure 6** (a) Summary of the EDX data and (b) and (c) EDX *spectra* of as-formed and crystalline (annealing in air at 500°C for 1 and 4 h) $TiO_2$ nanotube membranes of different thickness (5.0, 10 and 28 µm).

**Figure 7** Photocurrent transients of as-formed and crystalline (annealing in air at 500°C for 1 and 4 h) $TiO_2$ nanotube membranes (10 µm-thick layers) and compact oxides, measured at different wavelengths (in the visible region). The transients in (a) and (b) are measured using the IPCE setup (selecting the proper wavelength with the monochromator), while the transients in (c) and (d) are measured under laser light irradiation at 473 nm.

Figures

**Scheme 1**

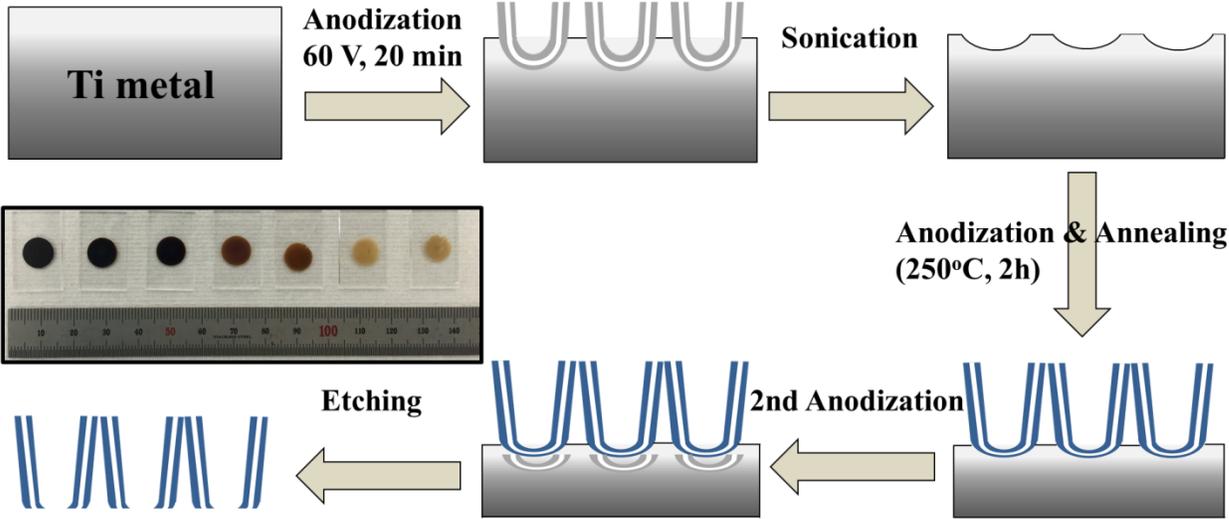

**Figure 1**

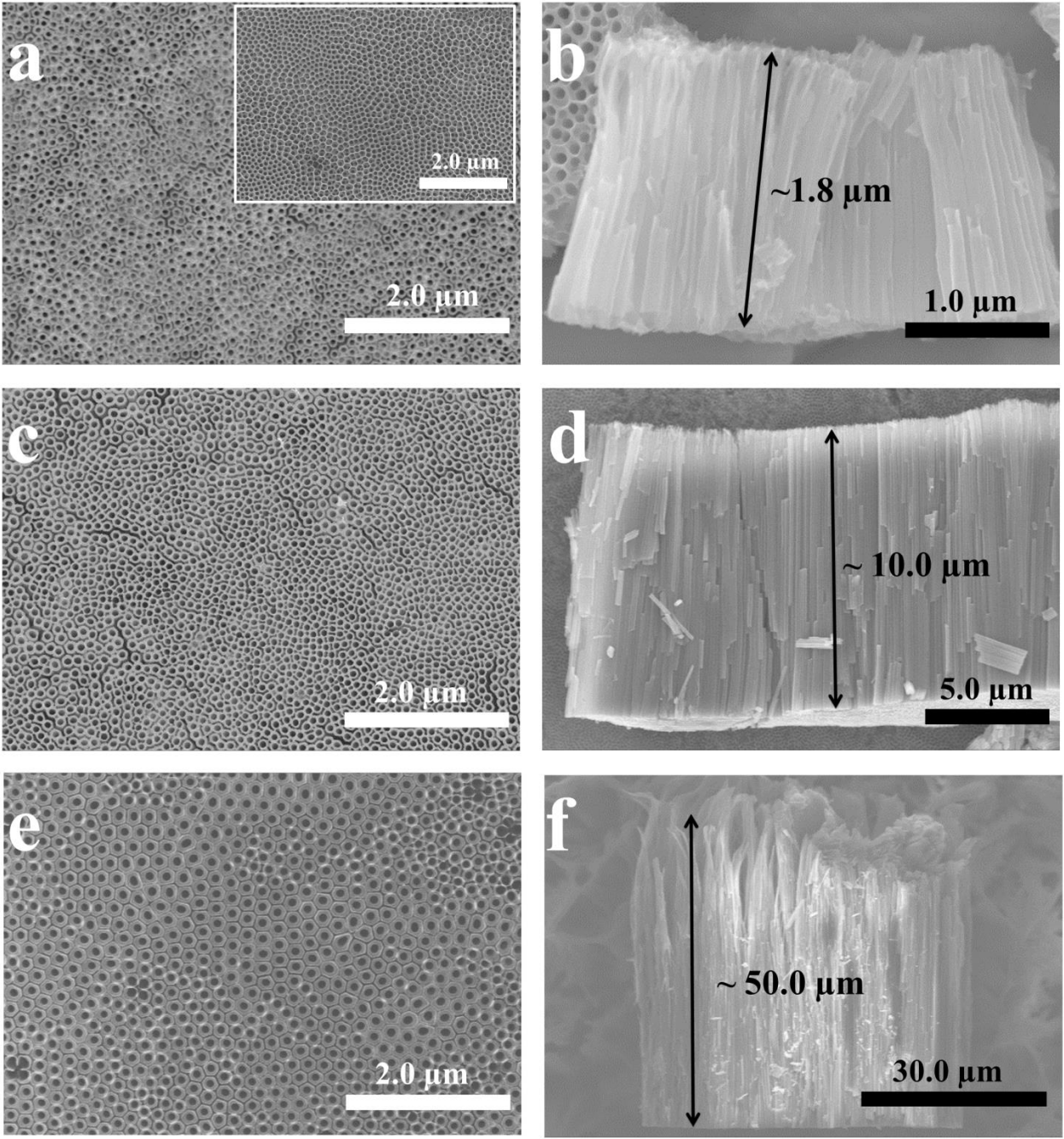

**Figure 2**

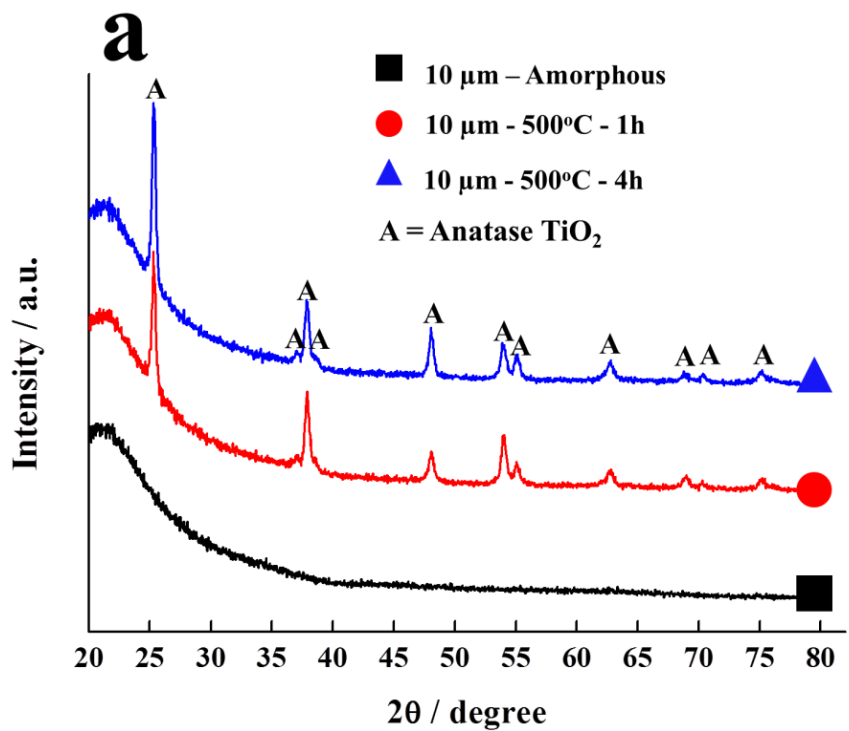

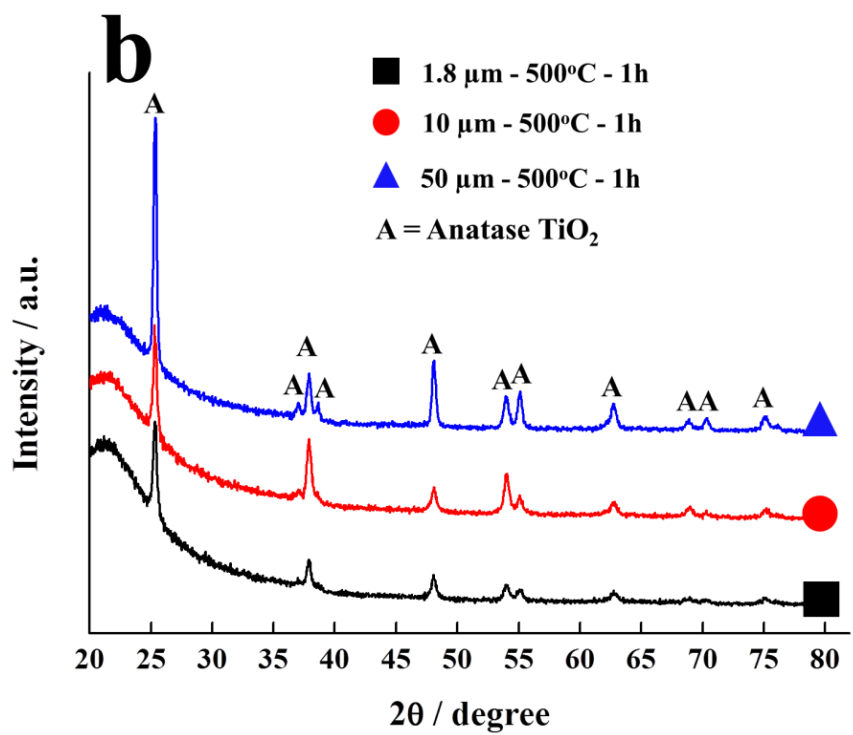

**Figure 3**

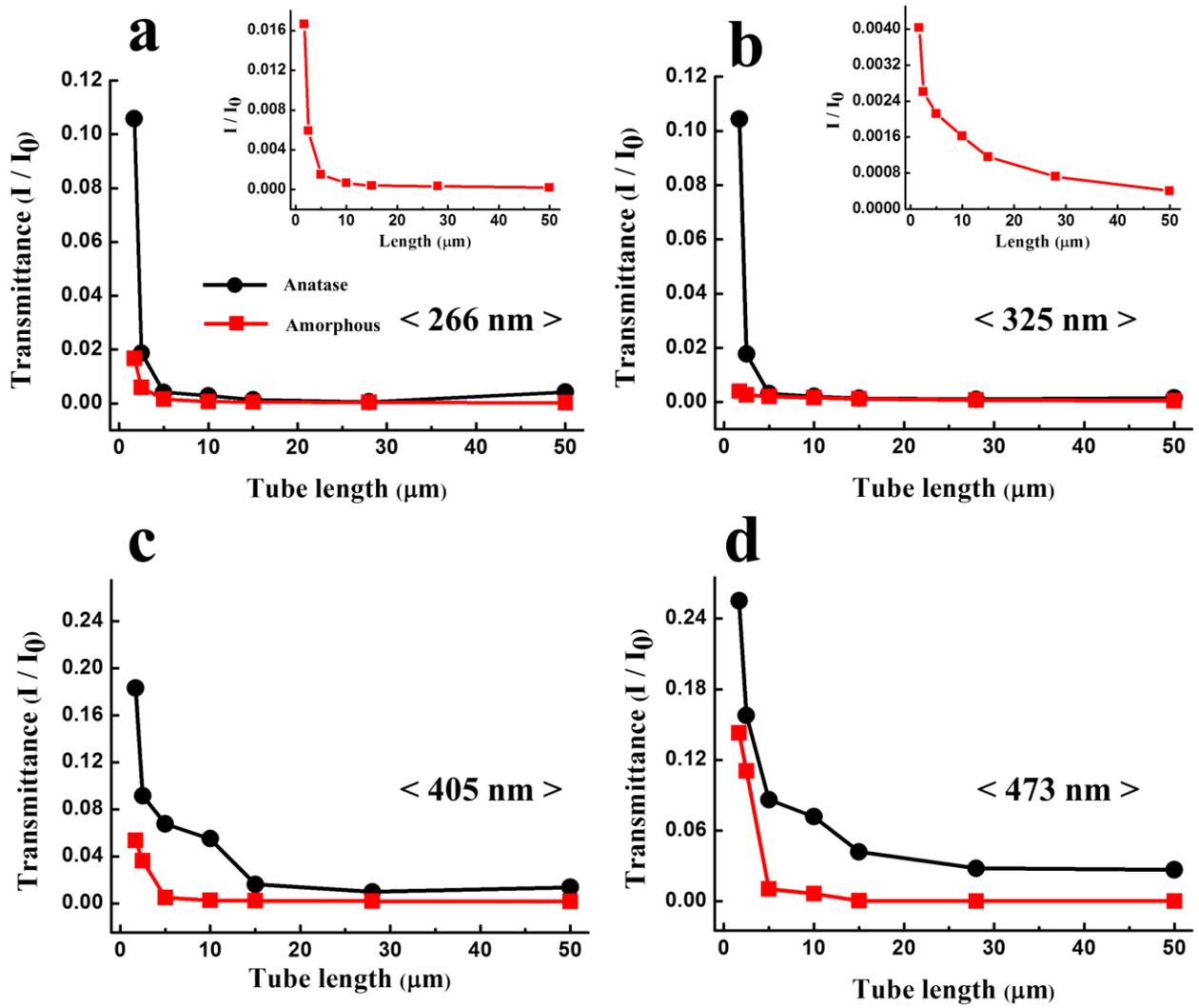

**Figure 4**

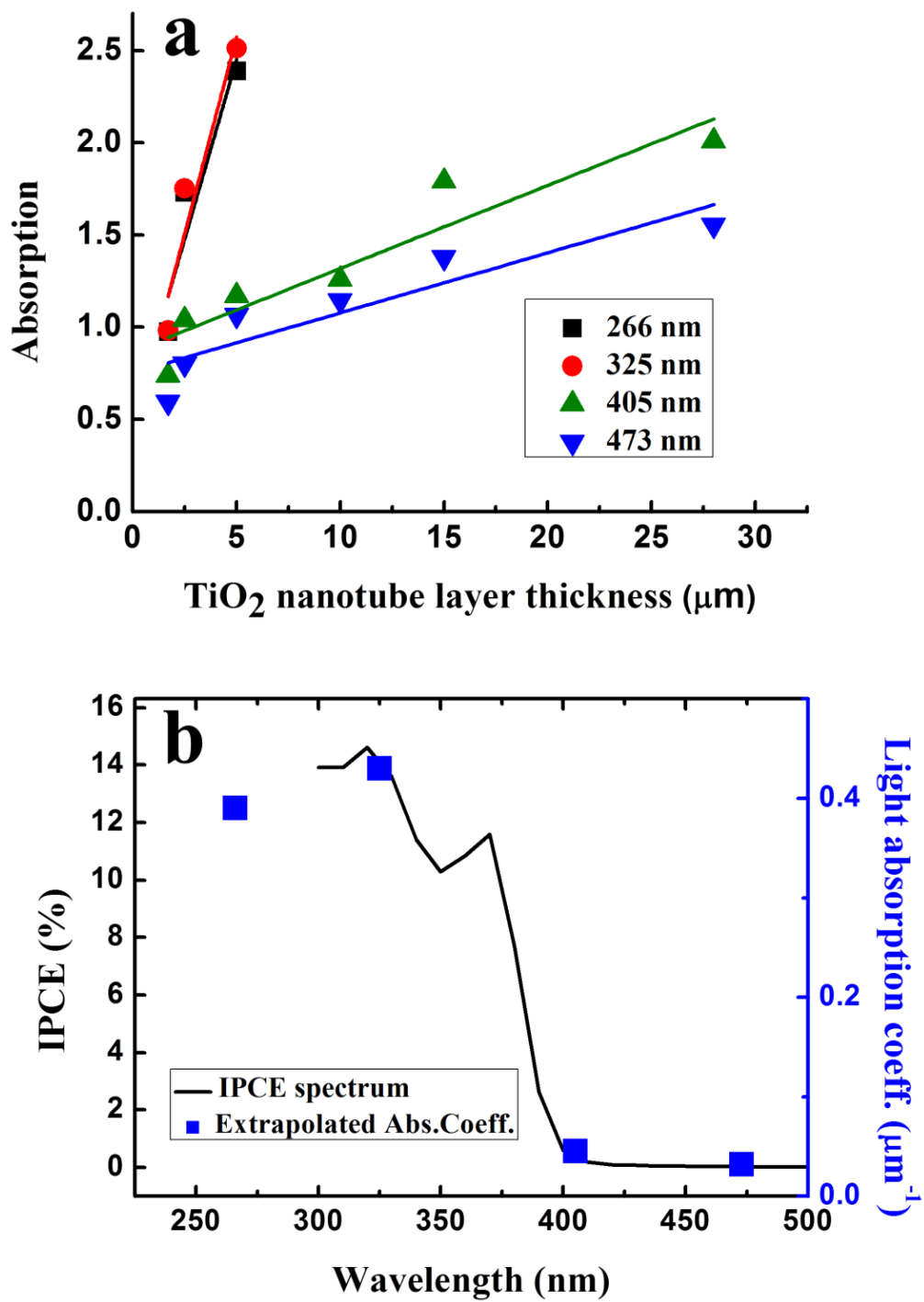

**Figure 5**

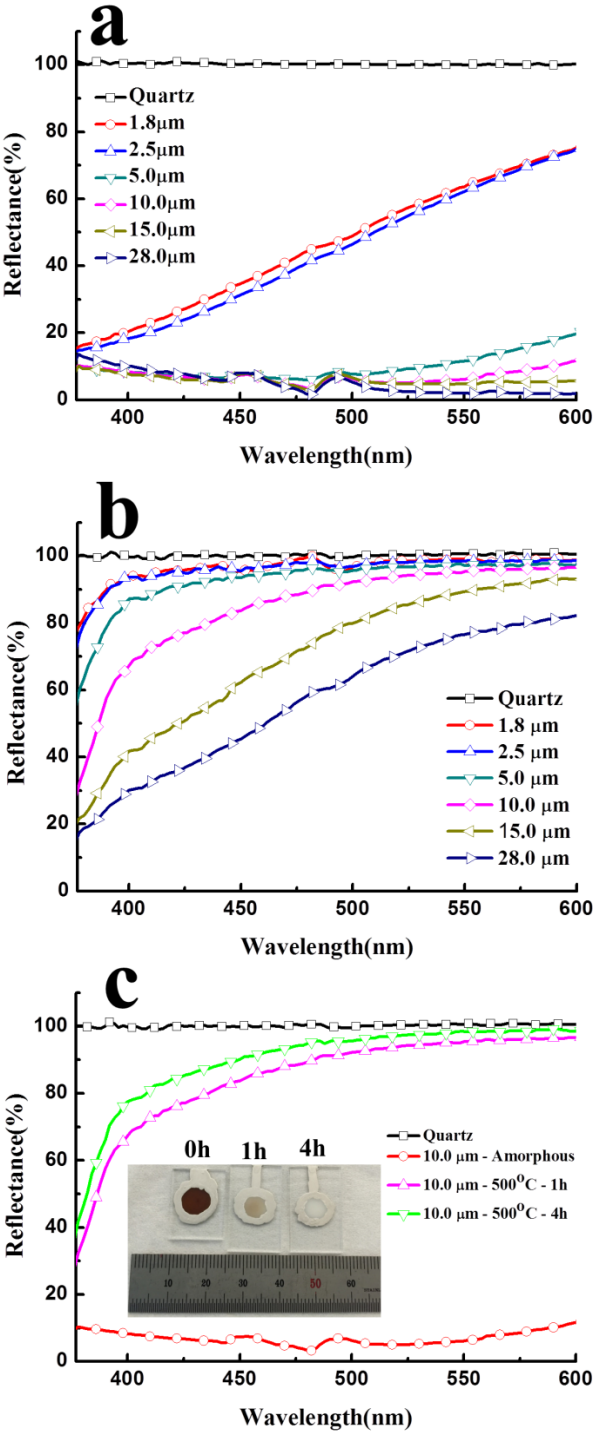

**Figure 6**

| | At% |
|---|---|
| 5.0 μm (500°C - 1h) | 3.06% |
| 10.0 μm (500°C - 1h) | 2.63% |
| 10.0 μm (Amorphous) | 10.41% |
| 28.0 μm (500°C - 1h) | 2.58% |

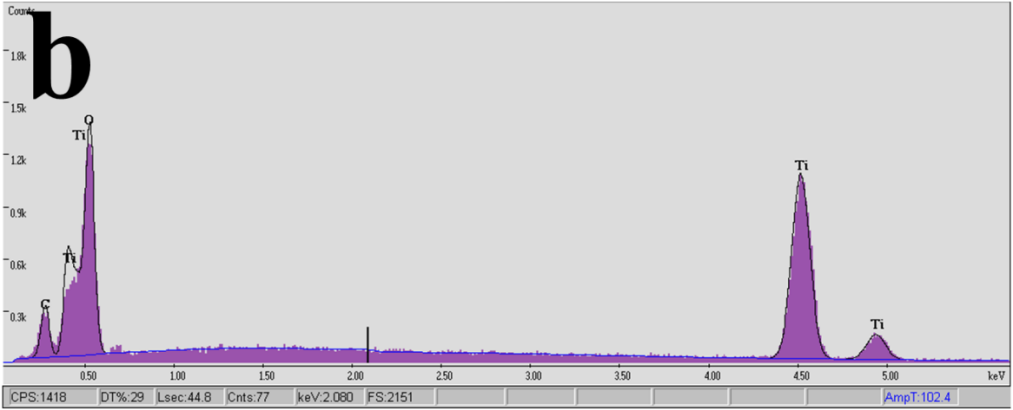

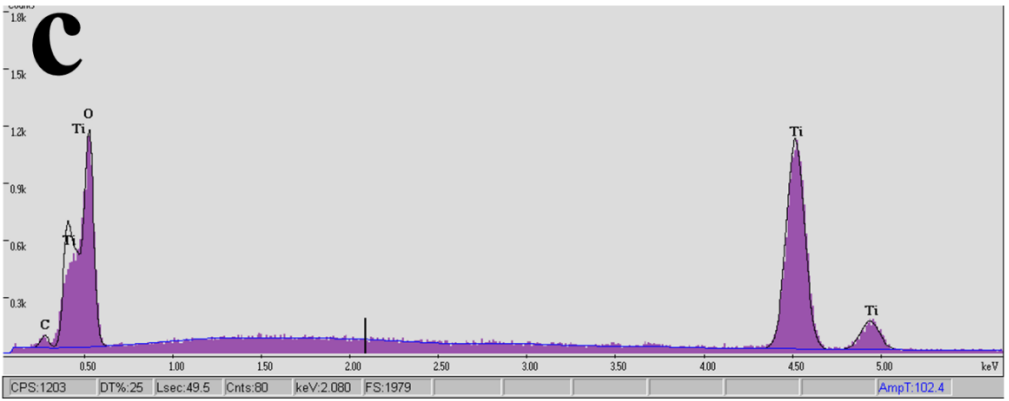

**Figure 7**

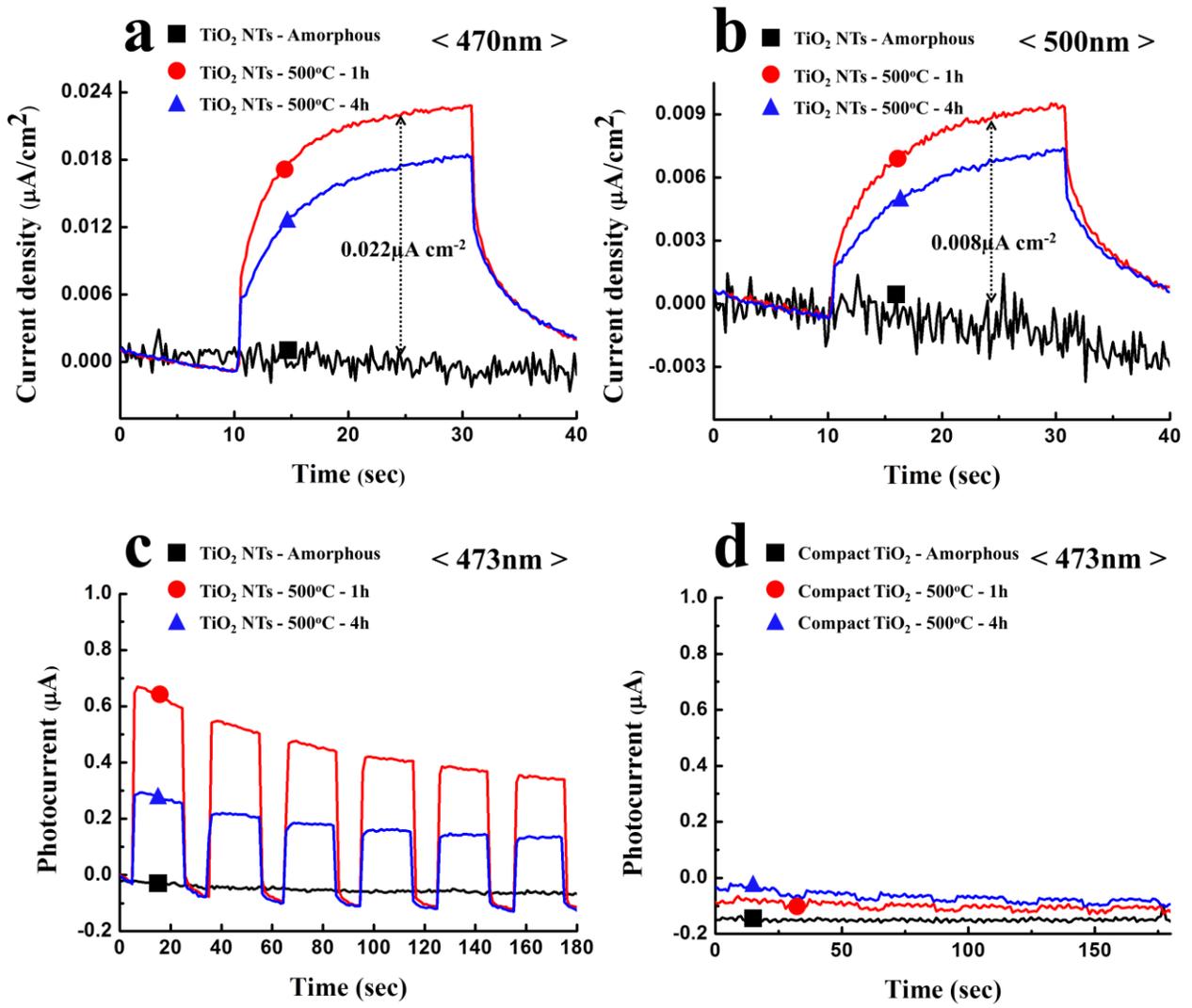